\documentclass[,final]{aipproc}

\layoutstyle{8x11double}

\newcommand{\bee}{\begin{eqnarray}}
\newcommand{\ene}{\end{eqnarray}}

\newcommand{\vxi}{\mbox{\boldmath{$\xi$}}}

\begin{document}

\title{Multimode Cepheids in the Large Magellanic Cloud - challenges for theory}


\author{W.A. Dziembowski}{
  address={Warsaw University Observatory, Al.~Ujazdowskie~4,
00-478~Warsaw, Poland}
 ,altaddress={Copernicus Astronomical Center, ul.~Bartycka~18, 00-716~Warsaw, Poland}
}

\author{R. Smolec}{
  address={Copernicus Astronomical Center, ul.~Bartycka~18, 00-716~Warsaw, Poland}
}

\begin{abstract}
Data on multimode Cepheids from OGLE-III catalog of the LMC Cepheids
are confronted with results from model calculations. Models whose
radial mode periods are consistent with observation are not always in
agreement with published evolutionary models. Nonradial mode
interpretation is considered for the cases of unusual period ratios.
The greatest challenge for stellar pulsation theory is explanation of
double-mode pulsators with period ratios near 0.6.
\end{abstract}

\maketitle

\section{Introduction}

The OGLE-III catalog of the LMC Cepheids \cite{Sosz08b} contains a
large number of objects with more than one mode excited. There are
61 classical beat Cepheids pulsating in the fundamental and first
overtone modes (F/1O) and 203 objects pulsating in the first
two overtones (1O/2O). The rare types of multimode radial
pulsators include two 1O/3O objects, two triple-mode pulsators
of the F/1O/2O type and three of the 1O/2O/3O type
\cite{Sosz08a}. Of these rare Cepheids, only two 1O/2O/3O objects
were known before \cite{Mos04}. More common (29 objects) are first
overtone pulsators with additional periods which are equal about 0.6 of the
dominant ones and cannot be explained in terms of radial
overtones. We denote them as the 1O/X pulsators. In Figure 1,
we show period distribution of various types of short period Cepheids.

In a number of LMC Cepheids, additional periods close to the
dominant modes are detected. These were found in a relatively small fraction
(4\%) of fundamental mode Cepheids but in a significant fraction of
objects with exited overtones (e.g. 18\% in 1O and 36\% in 1O/2O).

\begin{figure}
  \includegraphics[height=.23\textheight]{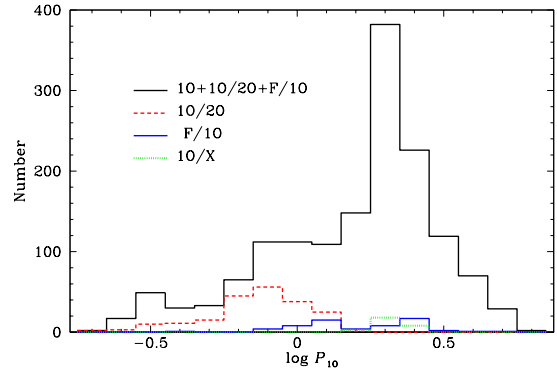}
  \caption{Period distribution for the LMC Cepheids with the
first overtone excited. The 1O/2O pulsators constitute 13.8\%
of the sample but dominate in a certain period range. The F/1O
pulsators constitute only 4.1\%. The 1O/X objects (1.9\%
of the sample)  and the long-period F/1O pulsators occur in the same
period range.}
\end{figure}

\section{Inference from data on two radial mode periods and $W_I$}

Most of multimode pulsators are found among short-period Cepheids,
which, as first
noted in \cite{alcock} and \cite{alibert}, are not well explained in terms of evolutionary models of
core-helium burning stars (second and third crossings of the
instability strip). Some of them may be in the first crossing phase,
which is much faster, but there are also problems with this
interpretation.

Data on two periods of radial modes and on Wesenheit indexes, such
as shown in Figures 2 and 3, yield strong constraints on stellar
models. In principle, color data may also be used but they are less
accurate. Instead, as a constrain on effective temperatures, a
requirement of mode instability was adopted in recent works
exploiting data on multimode Cepheids (\cite{mosad}, \cite{buchlerszabo}, \cite{buchler} and \cite {DaS},
hereafter DS). In DS, which was devoted almost
exclusively to the 1O/2O LMC Cepheids, the $W_I$ and period data
(Figure 3) were used to constrain parameters of these objects,
in particular, their masses.

In DS and here, model values of $W_I$ were obtained assuming
distance modulus to LMC 18.5\thinspace{}mag \cite{schaefer}. The
error of $\pm$0.05\thinspace{}mag translates to $\pm$7\%
uncertainty in stellar mass. Lines  shown in  Figures 2 and 3 were
obtained for models calculated with the OP opacity data \cite{seaton}
 for the new solar heavy element mixture \cite{asplund},
adopting $Z=0.006$. We carried linear
nonadiabatic calculations of radial modes for deep unfitted envelope
models within the relevant temperature range using the code
described in \cite{SiM}. For luminosity, we first
used the values obtained from evolutionary models in the first
crossing phase, calculated assuming no overshooting. Then, we
considered models with luminosity increments $\Delta\log L=0.2$ and
0.4 to reproduce the whole range of observed parameters. Models with
typical overshooting are brighter by $\Delta\log L\approx0.15$.
Those in the second and third crossings are brighter than models in
the first crossing by $\Delta\log L\approx0.25$ (see \cite{pietrinferni}).

\begin{figure}
  \includegraphics[height=.34\textheight]{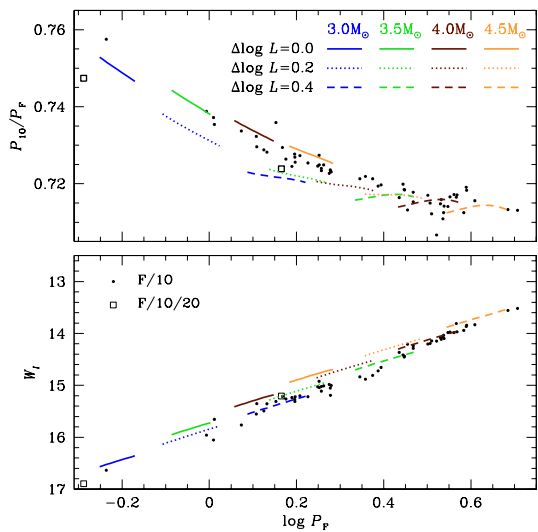}
  \caption{Data (points) on  the F/1O Cepheids. The Petersen diagram (top) and
PL relations (bottom) compared with values for selected models of
indicated mass (segments). The segments correspond to the
simultaneous instability of the F and 1O modes for models of
specified mass and luminosity. Luminosity increments $\Delta\log L$
refer to evolutionary models calculated with $\alpha_{\rm ov}=0$.}
\end{figure}

\begin{figure}
  \includegraphics[height=.34\textheight]{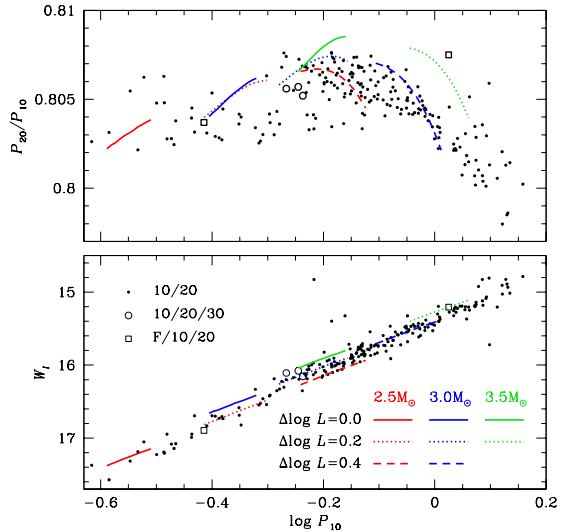}
  \caption{The same as Figure 2 but for the 1O/2O Cepheids.}
\end{figure}

In the lower panels of Figures 2 and 3, we may see that models of
constant mass form nearly a single line in the $\log P - W_I$ plane
for specified mode and metallicity. In fact, the dependence on
metallicity is quite weak. This means that, once we have credible
$W_I$ and the LMC distance, we have quite accurate assessment of a
Cepheid mass.

The upper panel of Figure 2 shows the classical Petersen diagram. We
may see that  the measured period ratios are well reproduced with
our models. Comparing plots in the two panels, we note certain
problem for $\log P_{\rm F}<0.3$. Models with masses near $3M_\odot$
are in agreement with the observational PL relation if some
luminosity excess (presumably due to overshooting) is allowed. Thus,
the objects may be understood as first crossers. The calculated
period ratio is somewhat low. The agreement would improve by
choosing lower $Z$ value (see \cite{buchler}). For the F/1O Cepheids
with $\log P_{\rm F}>0.3$, both the period ratios and the Wesenheit
indexes, point to masses near $4 M_\odot$ and the luminosity
excess $\Delta\log L\approx0.4$. These are presumably helium burning
objects.

Figure 3 is from DS, where it was concluded that great majority of the
LMC 1O/2O Cepheids  have masses $M=3.0\pm 0.5M_\odot$. The objects
with $\log P_{\rm 1O}< -0.2$ are well explained with post-main
sequence stellar models crossing the instability strip for the first
time, calculated assuming no or moderate overshooting. For those
with longer periods, which constitute the majority of the sample, a
significant luminosity excess is needed. Interpretation in terms of
first crossing models requires an overlarge overshooting, perhaps
connected with fast rotation. The difficulty of this explanation is
the short crossing time of the instability strip while the number of
the objects showing the luminosity excess is relatively high. If
these are helium burning objects then the difficulty is the low
inferred mass. Standard evolutionary tracks for stars with $M<3.5$
and acceptable metallicity do not enter the instability strip in
this evolutionary phase.

All five triple-mode Cepheids are most likely first crossing
objects. Detailed modeling of two 1O/2O/3O objects implied masses
near $3.3 M_\odot$ \cite{mosad}. The third one has very similar periods and
Wesenheit index, hence, must have a similar mass. Models of similar
mass and calculated with moderate overshooting well reproduce data
on the two 1O/3O Cepheids. The two F/1O/2O objects occur at very
different periods. The one at $\log P_{\rm 1O}=-0.4$ has mass $M<3.0
M_\odot$ and that at $\log P_{\rm 1O}=0.03$ has mass $M\approx3.5
M_\odot$.

\section{Nonradial modes and troublesome period ratios}

In the radiative interiors of evolved stars, the Brunt-V\"ais\"al\"a
frequency, $N$, is by orders of magnitude larger than frequencies,
$\omega$, of radial modes. This implies that all nonradial modes
propagate as high-order gravity waves with radial wave number,
$k_r$, satisfying $rk_r=\sqrt{\ell(\ell+1)}{N\over\omega}\gg1$. The
displacement amplitude is $\vxi\propto[C_+{\rm e}^{{\rm i}\Psi}+
C_-{\rm e}^{-{\rm i}\Psi}] {\rm e}^{-{\rm i}\omega t}$ with
$\Psi=\int dr k_r$.

Large $k_r$ leads to large radiative losses. Global modes exist as
long as $|C_-|\sim|C_+|$, and they are unstable if the driving
effects in outer layers are strong enough. However, the wave braking
occurring at $k_r|\vxi|\approx1$ may prevent growth of an unstable
mode amplitude to observable level. Once it occurs, the reflected
wave must be ignored ($|C_-|\approx0$) and this implies wave energy
losses. Only with a resonant energy transfer from a nearby radial
mode, detectable amplitude of nonradial modes may be maintained.
This mechanism, suggested as an explanation for  peaks close to
radial mode frequencies in RR Lyrae stars \cite{DaM}, may apply also 
to Cepheids.

Strong radiative damping in the interior leads to $|C_-|\ll|C_+|$.
This again justifies ignoring the reflected wave. At sufficiently
high-$\ell$, there are well-trapped modes, which, in spite of the
wave losses, remain unstable \cite{osaki}, \cite{D}. Due to
cancelation of opposite sign contributions, the net light changes
resulting from such mode excitation are expected low.

The $P_{\rm X}/P_{\rm 1O}$ period ratios are concentrated around 0.6
and 0.62. There are no systematic differences in the positions in
the PL diagram between the two groups of the 1O/X Cepheids. Both are
well inside the 1O band. No radial mode may account for such period
ratios. The closest but still far away is $P_{\rm 3O}/P_{\rm 1O}$,
as we may see in Figure 4.

Interpretation in terms of nonradial modes is also difficult. In
models constrained by observational data, the only modes within the
$P_{\rm X}$ range which do not suffer from strong damping in the
deep interior are F-modes (no nodes outside g-wave propagation zone)
of $\ell$ between 40 and 50. For modes of degrees $\ell>6$, the
light changes arise mainly from geometrical distortion of the
surface. Then, for even degree modes, the net light amplitude
decreases with $\ell$ only as $\ell^{-0.5}$. Still the amplitude
reduction is large and would require intrinsic pulsation amplitudes
of these modes significantly larger than those of the 1O modes.
Moreover, we have no explanation for the preference to 0.6 and 0.62 period ratios.

\begin{figure}
  \includegraphics[height=.33\textheight]{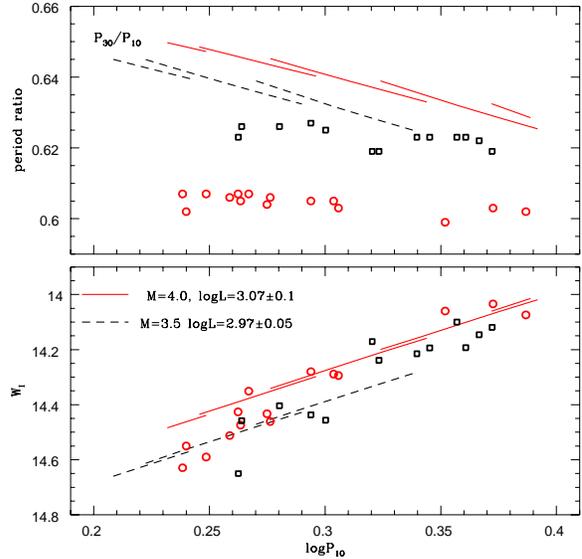}
  \caption{Period ratio (top) and Wesenheit  indexes (bottom) for 1O/X
Cepheids (symbols) compared with calculated values for the 1O/3O
Cepheid models (segments). Unfitted envelope models were calculated
for temperatures corresponding to the 1O instability range (3O is
stable in these models).}
\end{figure}

\begin{theacknowledgments}
  This work has been supported by the Polish MNiSW grant No 1 P03D 011 30.
\end{theacknowledgments}

\end{document}